\documentclass[aps]{revtex4-2}
\usepackage{graphicx}
\usepackage{bm}
\usepackage[utf8]{inputenc}
\usepackage[T1]{fontenc}
\usepackage{mathptmx}
\usepackage{etoolbox}

\begin{document}

\title{Interactions of fractional solitons with local defects: Stabilization
and scattering}
\author{Thawatchai Mayteevarunyoo$^1$}
\author{Boris A. Malomed$^{2,3}$}
\affiliation{$^1$Department of Electrical and Computer Engineering,	Faculty of
Engineering, Naresuan University, Phitsanulok 65000, Thailand}
\affiliation{$^2$Department of Physical Electronics, School of Electrical Engineering, 
Faculty of Engineering, Tel Aviv University, P.O.B. 39040, Ramat Aviv, Tel Aviv, Israel}
\affiliation{$^3$Instituto de Alta Investigaci\'{o}n, Universidad de
Tarapac\'{a}, Casilla 7D, Arica, Chile}

\begin{abstract}
Stability is an essential problem in theoretical and experimental studies of
solitons in nonlinear media with fractional diffraction, which is
represented by the Riesz derivative with L\'{e}vy index (LI) $\alpha $,
taking values $\alpha <2$. Fractional solitons are unstable at $\alpha \leq
1 $ or $\alpha \leq 2$ in uniform one-dimensional media with the cubic or
quintic self-focusing, respectively. We demonstrate that, in these cases,
the solitons may be effectively stabilized by pinning to a delta-functional
trapping potential (attractive defect), which is a relevant setting in
optical waveguides with the effective fractional diffraction. Using the
respective fractional nonlinear Schr\"{o}dinger equation with the
delta-functional potential term, we find that, in the case of the cubic
self-focusing, the fractional solitons are fully stabilized by the pinning
to the defect for $\alpha =1$, and partly stabilized for $\alpha <1$. In the
case of the quintic self-focusing, the full and partial stabilization are
found for $\alpha =2$ and $\alpha <2$, respectively. In both cases, the
instability boundary is exactly predicted by the Vakhitov-Kolokolov
criterion. Unstable solitons spontaneously transform into oscillating
breathers. A variational approximation (VA) is elaborated parallel the
numerical analysis, with a conclusion that the VA produces accurate results
for lower LI values, i.e., stronger fractionality. In the cubic medium,
collisions of traveling stable solitons with repulsive and attractive
defects are addressed too, demonstrating outcomes in the form of rebound,
splitting, and passage.
\end{abstract}

\maketitle


\textbf{The consideration of linear and nonlinear systems with effective
fractional diffraction and/or dispersion is one of basic directions in the
development and investigation of complexity in various fields of physics,
such as optics, quantum matter, solid state, classical field theory, and
others. While the fractional calculus was known, as a mathematical
curiosity, in the course of the past 200 years, the concept of fractional
dispersion had been introduced in physics, in the form of fractional quantum
mechanics, by N. Laskin, in 2000. In this context, the commonly known
one-dimensional (1D) operator of the kinetic energy, }$-d^{2}/dx^{2}$\textbf{%
, is replaced by the fractional-order \textit{Riesz derivative}, }$\left(
-d^{2}/dx^{2}\right) ^{\alpha /2}$\textbf{, with the \textit{L\'{e}vy index }%
(LI) }$\alpha $\textbf{, so that the canonical (non-fractional) theory
corresponds to }$\alpha =2$\textbf{, while the fractality corresponds to }$%
\alpha <2$\textbf{. With all its theoretical appeal, fractional quantum
mechanics remains far from experimental realization. To advance the
fractality concept to other branches of physics, and push it closer to a
possibility of experimental implementation, it was proposed by S. Longhi in
2015, using the commonly known similarity between the quantum-mechanical Schr%
\"{o}dinger equation and the parabolic equation for the classical paraxial
propagation (diffraction) of light, to emulate fractional quantum mechanics
by effective fractional diffraction in optical cavities. Only in 2023, the
first experimental realization of fractional optics was reported, in terms
of the light transmission in optical fiber cavities which feature effective
fractional dispersion. While the first theoretical and experimental
implementations of the fractional light propagation were reported in the
linear form, the intrinsic nonlinearity of optical materials suggest a
possibility of consideration of nonlinear modes, such as fractional
solitons, supported by the interplay of the fractional
diffraction/dispersion and material nonlinearity. This possibility was
recently elaborated in many theoretical works. The first experimental
demonstration of fractional temporal solitons in fiber cavities was reported
very recently \cite{Shilong-nonlinear,Martijn}. In this context, the
stability of fractional solitons is a crucially important issue. In
particular, in the 1D space, the interplay of the usual Kerr (cubic)
self-focusing with the fractional diffraction gives rise to stable solitons
in the interval of }$1<\alpha \leq 2$\textbf{, while they are unstable,
because of the occurrence of the \textit{critical} and \textit{supercritical}
collapse, at }$\alpha =1$\textbf{\ and }$\alpha <1$\textbf{, respectively.
In the case of the quintic self-focusing, which is also possible in optical
materials, fractional solitons are unstable at all values of }$\alpha \leq 2$%
\textbf{. Therefore, stabilization of fractional solitons is an issue of
major significance. The present paper demonstrates that the stabilization
may be provided by using a linear potential in the form of a local potential
well (attractive local defect). By means of a systematic numerical
investigation and an analytical variational approximation, combined with the
well-known \textit{Vakhitov-Kolokolov} \textit{criterion}, it is
demonstrated that, in the case of the cubic self-focusing, the fractional
solitons, pinned to the attractive defect, are completely and partly
stabilized for }$\alpha =1$\textbf{\ and }$\alpha <1$\textbf{, respectively.
Similarly, in the case of the quintic self-focusing, the complete and
partial stabilization of the pinned solitons is demonstrated for }$\alpha =2$%
\textbf{\ and }$\alpha <2$\textbf{, respectively.}

\section{Introduction and the model}

In the course of the past two decades, models based on linear and nonlinear
Schr\"{o}dinger equations with fractional dispersion, and similar equations
with fractional operators, have drawn much interest in physical kinetics
\cite{Za1,Za2}, quantum theory \cite{Laskin1,Stickler,Pinsker,Laskin-book},
and optics \cite{Longhi}, see also reviews \cite{Photonics,Chaos}. The first
experimental realization of the linear Schr\"{o}dinger equation with the
effective fractional group-velocity dispersion was performed in fiber-laser
cavities operating in the linear regime \cite{Shilong-linear}. The interplay
of the fractional dispersion and intrinsic nonlinearity of optical materials
gives rise to fractional nonlinear Schr\"{o}dinger equations (FNLSEs), which
produce various solutions in the form of fractional solitons \cite%
{frac-sol,LDnonlocal,LZcoupled}.Very recently, fiber optics was employed
to demonstrate the creation of such solitons in the temporal domain \cite%
{Shilong-nonlinear,Martijn}.

The one-dimensional (1D) FNLSE is introduced, in the scaled form, as%
\begin{equation}
i\frac{\partial \psi }{\partial z}=\frac{1}{2}\,\left( -\frac{\partial ^{2}}{
\partial x^{2}}\right) ^{\alpha /2}\psi -N(|\psi |^{2})\!\!\psi +U(x)\psi ,
\label{general}
\end{equation}%
It is written in the notation adopted for the propagation of optical waves,
with complex amplitude $\psi \left( x,z\right) $, in the spatial domain
spanned by the propagation distance $z$ and transverse coordinate $x$. The
equation may include an effective real potential $U(x)\sim -\delta n(x)$,
where $\delta n(x)$ is a local change of the refractive index. Term $N(|\psi
|^{2})\!\!\psi $ in Eq. (\ref{general}), with $N(|\psi |^{2})\!\!>0$,
represents the self-focusing nonlinearity, in the general form

The fractional-diffraction operator in Eq. (\ref{general}), with L\'{e}%
vy-index (LI) $\alpha \leq 2$ \cite{Mandelbrot}, is represented by the Riesz
derivative \cite{Riesz}. Unlike abstract concepts of fractional derivatives,
such as the Caputo operator \cite{Caputo}, the definition of the Riesz
derivative is based on the natural idea that, in the Fourier space with
wavenumber $p$, which is dual to coordinate $x$, the fractional
differentiation amounts to the multiplication by $|p|^{\alpha }$. This idea
leads to the definition based on the cascaded direct and inverse Fourier
transforms:%
\begin{equation}
\left( -\frac{\partial ^{2}}{\partial x^{2}}\right) ^{\alpha /2}\ \psi
\equiv D_{\text{Riesz}}^{(\alpha )}=\frac{1}{2\pi }\int_{-\infty }^{+\infty
}dp|p|^{\alpha }\int_{-\infty }^{+\infty }d\tilde{x}\exp \left( ik\left( x-%
\tilde{x}\right) \right) \psi \left( \tilde{x}\right) .  \label{R}
\end{equation}%
Thus, the Riesz derivative is actually an integral operator, rather than a
differential one. In the limit case of $\alpha =2$ it carries over to the
usual second derivative, $D_{\text{Riesz}}^{(\alpha =2)}=-\frac{\partial ^{2}%
}{\partial x^{2}}$.

In the free space ($U(x)=0$), the self-focusing nonlinearity in Eq. (\ref%
{general}) gives rise to a family of fractional solitons with real
propagation constant $k>0$ and real function $u(x)$,
\begin{equation}
\psi \left( x,z\right) =e^{ikz}u(x).  \label{psiphi}
\end{equation}%
The soliton is characterized by its power,%
\begin{equation}
P=\int_{-\infty }^{+\infty }\left[ u(x)\right] ^{2}dx.  \label{power}
\end{equation}%
In the case of the power-law nonlinear term,
\begin{equation}
N\left( |\psi |^{2}\right) =\left\vert \psi \right\vert ^{2\sigma }\!\!,
\label{nonlin}
\end{equation}%
the scaling properties of Eq. (\ref{general}) in the free space ($U(x)=0$)
predict the following dependence between the propagation constant and power:%
\begin{equation}
P\sim k^{\left( \alpha -\sigma \right) /\left( \alpha \sigma \right) }.
\label{kP}
\end{equation}

The soliton solutions of the non-fractional NLSE ($\alpha =2$) with the
general form (\ref{nonlin}) of the power-law nonlinearity can be easily
found in the exact form,%
\begin{equation}
u(x)=\left[ \sqrt{\left( \sigma +1\right) k}\mathrm{sech}\left( \sigma \sqrt{%
2k}x\right) \right] ^{1/\sigma }.  \label{sol}
\end{equation}%
The integral power (\ref{power}) of solitons (\ref{sol}) is%
\begin{equation}
P_{\alpha =2,\sigma }(k)=\sqrt{\frac{\pi }{2}}\frac{\left( \sigma +1\right)
^{1/\sigma }\Gamma \left( 1/\sigma \right) }{\sigma \Gamma \left( \left(
2+\sigma \right) /\left( 2\sigma \right) \right) }k^{\left( 2-\sigma \right)
/\left( 2\sigma \right) }  \label{P(k)}
\end{equation}%
\cite{Beijing}, cf. Eq. (\ref{kP}), where $\Gamma $ is the Euler's
Gamma-function.

If LI is not too low, \textit{viz}., $\alpha >\sigma $, relations (\ref{kP})
and (\ref{P(k)}) (the latter one corresponds to $\alpha =2$) satisfy the
celebrated Vakhitov-Kolokolov (VK) criterion \cite{Vakh,Berge,Fibich}, $%
dP/dk>0$, which is the necessary (although not sufficient) stability
condition of the soliton family. In the case of $\alpha \leq \sigma $, when
the family does not satisfy the VK criterion, the instability may initiate
the onset of the \textit{critical} and \textit{supercritical} collapse
(spontaneous blowup leading to the emergence of singular states) in the
cases of $\alpha =\sigma $ and $\alpha <\sigma $, respectively \cite%
{frac-coll,Chaos}. In the former case, the corresponding family of unstable
\textit{Townes solitons} (TSs) is degenerate, in the sense that its power
takes a single value which does not depend on $k$, in accordance with Eq. (%
\ref{kP}). In particular, for the non-fractional ($\alpha =2$) NLSE with the
quintic self-focusing term ($\sigma =2$ in Eq. (\ref{nonlin})), Eq. (\ref%
{sol}) produces the commonly known family of 1D TS solutions \cite{AS}. The
single value of power (\ref{power}) for this TS family is given by Eq. (\ref%
{P(k)}) with $\alpha =0$:%
\begin{equation}
P_{\mathrm{TS}}^{\left( \alpha =\sigma =2\right) }=\sqrt{\frac{3}{2}}\frac{%
\pi }{2}\approx 1.92.  \label{PTS}
\end{equation}

For the realization is optics, most relevant is the Kerr (cubic)
nonlinearity, corresponding to $\sigma =1$ in Eq. (\ref{nonlin}). The cases
of the quintic and septimal nonlinearities, which correspond to $\sigma =2$
and $3$, respectively, are relevant too, as they can be experimentally
realized in the light propagation in colloidal suspensions of metallic
nanoparticles \cite{Cid1,Cid2}.

In this connection, it is relevant to mention that LI values $\alpha \leq 1$
were not considered in a majority of theoretical studies dealing with FNLSEs
precisely for the reason that they give rise to the collapse in the
combination with the Kerr nonlinearity. Nevertheless, the methods proposed
for the creation of the effective fractional diffraction \cite{Longhi} and
dispersion \cite{Shilong-linear} are as appropriate for $\alpha \leq 1$ as
they are for $\alpha >1$.

Thus, the stability of fractional solitons is an essential problem, which
has drawn considerable interest. In particular, a possibility for the
creation and stabilization of fractional solitons under the action of
spatially periodic (lattice) potentials was elaborated in various forms

A possibility to create stable solitons in the 1D medium with the usual
(non-fractional) diffraction, $\alpha =2$, and the self-focusing
nonlinearity (\ref{nonlin}) with $\sigma \geq 2$, when the solitons in the
uniform medium are unstable, according to the VK criterion, was proposed in
Ref. \cite{Beijing}, by means of the delta-functional attractive potential,
which corresponds to%
\begin{equation}
U(x)=-\varepsilon \delta (x),  \label{eps}
\end{equation}%
in Eq. (\ref{general}), with $\varepsilon >0$. In optics, this potential
represents a narrow strip in the planar waveguide with a large refractive
index.

Exact solutions for solitons pinned to the delta-functional attractive
center (alias \textit{attractive defect}) are given by Eq. (\ref{sol}), with
replacement%
\begin{equation}
x\rightarrow \left( |x|+\Xi \right) ,~\Xi =\left( 2\sigma \sqrt{2k}\right)
^{-1}\ln \left( \frac{\sqrt{2k}+\varepsilon }{\sqrt{2k}-\varepsilon }\right)
.  \label{xi}
\end{equation}%
As seen from Eq. (\ref{xi}), the pinned solitons exist with values of the
propagation constant $k\geq k_{\min }=\varepsilon ^{2}/2$ ($k_{\min }$ is
the eigenvalue for the bound state produced by the linear equation with the
non-fractional diffraction, $\alpha =2$). Both the VK criterion and full
numerical analysis \cite{Beijing} demonstrate that the pinned solitons are
fully stable in the semi-infinite interval of $k_{\min }\leq k<\infty $, for
the critical quintic nonlinearity, $\sigma =2$, and in a finite interval, $%
k_{\min }\leq k\leq k_{\max }$, for $\sigma >2$. In particular, $k_{\max
}\approx 8\varepsilon ^{2}/\left[ \pi \left( \sigma -2\right) \right] ^{2}$
for $0<\sigma -2\ll 1$ \cite{Beijing}.

The objective of this work is to introduce the FNLSE model with the
nonlinearity (\ref{nonlin}) and delta-functional potential (\ref{eps}):

\begin{equation}
i\frac{\partial \psi }{\partial z}=\frac{1}{2}\,\left( -\frac{\partial ^{2}}{%
\partial x^{2}}\right) ^{\alpha /2}\psi -\left\vert \psi \right\vert
^{2\sigma }\!\!\psi -\varepsilon \delta (x)\psi ,  \label{NLS}
\end{equation}%
and analyze a possibility of the stabilization of fractional solitons pinned
to the attractive defect in the relevant cases, corresponding to $\alpha
\leq 1$ with the cubic self-focusing ($\sigma =1$), and $\alpha <2$ with the
quintic self-focusing ($\sigma =2$), when the solitons are completely
unstable in the free space ($U=0$). As mentioned above, these settings can
be implemented in fractional nonlinear optical waveguides, and may find
applications in various soliton-based data-processing schemes \cite{Turitsyn}%
.

The substitution of ansatz (\ref{psiphi}) in Eq. (\ref{NLS}) leads to the
real stationary equation:%
\begin{equation}
ku+\frac{1}{2}\,\left( -\frac{d^{2}}{dx^{2}}\right) ^{\alpha /2}u-u^{2\sigma
+1}-\varepsilon \delta (x)u=0.  \label{u}
\end{equation}%
By means of rescaling, we set $\varepsilon \equiv 1$ in Eqs. (\ref{NLS}) and
(\ref{u}) for $\alpha \neq 1$, while $\varepsilon $ should be kept as a free
parameter in the special case of $\alpha =1$.

The subsequent presentation is arranged as follows. Making use of the
possibility to represent Eq. (\ref{u}) in the Lagrangian form \cite%
{VA,frac-coll,Chaos}, the variational approximation (VA) for the fractional
solitons pinned to the attractive defect is developed in Section 2.
Numerical results for families of the pinned solitons, including the
analysis of their stability, are collected and compared to predictions of VA
in Section 3. In addition, collisions of traveling stable fractional
solitons with the local defect, attractive or repulsive (the latter one
corresponds to $\varepsilon <0$ in Eq. (\ref{eps})), in the case of $\sigma
=1$ (the cubic self-focusing) and $\alpha >1$, are considered in Section 4.
In the same section, we briefly consider symmetric two-soliton bound states
attached to the repulsive defect, concluding that such states exist but are
unstable against spontaneous symmetry-breaking splitting. The paper is
concluded by Section 5.

\section{The variational approximation (VA)}

The variational and numerical results are produced below with the ideal
delta-function in Eqs. (\ref{NLS}) and (\ref{u}) replaced by its standard
Gaussian regularization,
\begin{equation}
\tilde{\delta}(x)=\left( \sqrt{\pi }\xi \right) ^{-1}\exp \left( -\frac{%
x^{2} }{\xi ^{2}}\right) .  \label{delta}
\end{equation}%
The delta-function limit (\ref{eps}) for the potential is valid if the
regularization scale $\xi $ in approximation (\ref{delta}) is much smaller
than the inner scale of the soliton, which may be estimated as $\sim
k^{-1/\alpha }$ (cf. Eq. (\ref{kP})), i.e., under the condition of $\xi \ll
k^{-1/\alpha }$. The numerical results are presented below for $\xi =0.05$,
which satisfies the latter condition for relevant cases.

Taking into regard definitions (\ref{R}) of the Riesz derivative and
regularized delta-function (\ref{delta}), Eq. (\ref{u}) can be derived from
the following Lagrangian:%
\begin{eqnarray}
L &=&\frac{k}{2}\int_{-\infty }^{+\infty }dx\left[ u(x)\right] ^{2}+\frac{1}{%
8\pi }\int_{-\infty }^{+\infty }dp|p|^{\alpha }\int_{-\infty }^{+\infty
}dx\int_{-\infty }^{+\infty }d\tilde{x}\exp \left( ip\left( x-\tilde{x}%
\right) \right) u\left( x\right) u\left( \tilde{x}\right)   \nonumber \\
&&-\frac{1}{2\left( \sigma +1\right) }\int_{-\infty }^{+\infty }dx\left[ u(x)%
\right] ^{2\left( \sigma +1\right) }-\frac{\varepsilon }{2\sqrt{\pi }\xi }%
\int_{-\infty }^{+\infty }dx\exp \left( -\frac{x^{2}}{\xi ^{2}}\right) \left[
u(x)\right] ^{2}.  \label{L}
\end{eqnarray}%
An analytically tractable version of VA is based on the Gaussian ansatz with
amplitude $A$ and width $W$:%
\begin{equation}
u(x)=A\exp \left( -\frac{x^{2}}{2W^{2}}\right) ,  \label{ans}
\end{equation}%
the respective power (\ref{P}) being%
\begin{equation}
P_{\mathrm{\ VA}}=\sqrt{\pi }A^{2}W.  \label{Nans}
\end{equation}

The VA\ Lagrangian in the free space ($\varepsilon =0$) for $\sigma =1$,
produced by the substitution of ansatz (\ref{ans}) in Lagrangian (\ref{L}),
was known previously \cite{Chaos}. In the present case, which includes the
potential term $\sim \varepsilon $ and the general value of the nonlinearity
degree $\sigma $, the result is%
\begin{equation}
L_{\mathrm{\ VA}}=\frac{k}{2}P_{\mathrm{\ VA}}+\frac{\Gamma \left( \left(
1+\alpha \right) /2\right) }{4\sqrt{\pi }}\frac{P_{\mathrm{\ VA}}}{W^{\alpha
}}-\frac{\left( P_{\mathrm{\ VA}}\right) ^{\sigma +1}}{2\pi ^{\sigma
/2}\left( \sigma +1\right) ^{3/2}W^{\sigma }}-\frac{\varepsilon P_{\mathrm{\
VA}}}{2\sqrt{\pi }\sqrt{W^{2}+\xi ^{2}}},  \label{LVA}
\end{equation}%
where $\Gamma $ is the Gamma-function, and Eq. (\ref{Nans}) was used to
eliminate $A^{2}$ in favor of $P$. Note that the last term in Eq. (\ref{LVA}%
), which accounts for the effect of the attractive defect, carries over into
$\varepsilon P_{\mathrm{\ VA}}/\left( 2\sqrt{\pi }W\right) $ in the limit of
$\xi \rightarrow 0$, which corresponds to the limit of the ideal
delta-function in Eq. (\ref{delta}).

The variational (Euler-Lagrange) equations following from Lagrangian (\ref%
{LVA}) are%
\begin{equation}
\partial L_{\mathrm{\ VA}}/\partial P_{\mathrm{\ VA}}=\partial L_{\mathrm{\
VA}}/\partial W=0.  \label{EL}
\end{equation}%
These equations make it possible to find, in the framework of VA, values of
the propagation constant $k$ and width $W$ corresponding to given power $P$.
Thus, the first equation in system (\ref{EL}) yields an expression for $k$
in terms of $P$ and $W$:%
\begin{equation}
k=-\frac{\Gamma \left( \left( 1+\alpha \right) /2\right) }{2\sqrt{\pi }
W^{\alpha }}+\frac{\left( P_{\mathrm{\ VA}}\right) ^{\sigma }}{\pi ^{\sigma
/2}\sqrt{\sigma +1}W^{\sigma }}+\frac{\varepsilon }{\sqrt{\pi }\sqrt{
W^{2}+\xi ^{2}}}.  \label{k(P)}
\end{equation}%
The second equation in system (\ref{EL}) gives rise to a relation which
determines $W$ for given $P_{\mathrm{\ VA}}$:%
\begin{equation}
\left( P_{\mathrm{\ VA}}\right) ^{\sigma }=\frac{\left( \sigma +1\right)
^{3/2}}{\sigma }\pi ^{(\sigma -1)/2}\left[ \frac{\alpha }{2}\Gamma \left(
\frac{\alpha +1}{2}\right) W^{\sigma -\alpha }-\varepsilon \frac{W^{\sigma
+2}}{\left( W^{2}+\xi ^{2}\right) ^{3/2}}\right] .  \label{P}
\end{equation}

In the absence of the attractive defect ($\varepsilon =0$), Eq. (\ref{P})
demonstrates that, in the case of $\alpha =\sigma $, which gives rise to the
TS family, the corresponding VA-predicted single value of the power (one
that does not depend on $k$) is
\begin{equation}
\left( P_{\mathrm{\ VA}}\right) _{\mathrm{TS}}^{\left( \alpha =\sigma
,\varepsilon =0\right) }=\frac{1}{2}\left( \sigma +1\right) ^{3/2}\pi
^{(\sigma -1)/2}\Gamma \left( \frac{\sigma +1}{2}\right) .  \label{const}
\end{equation}%
In particular, Eq. (\ref{const}) yields%
\begin{equation}
\left( P_{\mathrm{\ VA}}\right) _{\mathrm{TS}}^{\left( \alpha =\sigma
=1,\varepsilon =0\right) }=\sqrt{2},~\left( P_{\mathrm{\ VA}}\right) _{%
\mathrm{TS}}^{\left( \alpha =\sigma =2,\varepsilon =0\right) }=\frac{1}{2}%
\sqrt{3\sqrt{3}\pi }\approx 2.02.  \label{VA-TS}
\end{equation}%
The comparison of the VA prediction, given by Eq. (\ref{VA-TS}) for the
power of the TS family with $\alpha =\sigma =2$, and the exact value of the
same power, as given by Eq. (\ref{PTS}), demonstrates that the accuracy of
the VA prediction is $\approx 5\%$.

As said above, the objective of the analysis is the possibility to stabilize
the fractional soliton by the pinning it to the attractive defect, with $%
\varepsilon >0$. This possibility can be further considered in the
analytical form for the limit case of $\xi =0$ (the ideal delta-function in
Eq. (\ref{u})), as the case of finite $\xi $ is too cumbersome for the
analytical study. Then, Eq. (\ref{P}) is simplified to%
\begin{equation}
\left( P_{\mathrm{\ VA}}\right) ^{\sigma }=\frac{\left( \sigma +1\right)
^{3/2}}{\sigma }\pi ^{(\sigma -1)/2}\left[ \frac{\alpha }{2}\Gamma \left(
\frac{\alpha +1}{2}\right) W^{\sigma -\alpha }-\varepsilon W^{\sigma -1}%
\right] .  \label{P2}
\end{equation}

The numerical results presented below in Fig. \ref{fig1} suggest that the
soliton pinned to the attractive defect exists with the power taking values
below a certain maximum, which corresponds to $dP_{\mathrm{\ VA}}/dW=0$, in
terms of the VA. As it follows from Eq. (\ref{P2}), the latter condition
yields no result for $\sigma =1$, while for $\sigma =2$ the maximum is
attained at $W=W_{m}$, with \
\begin{equation}
W_{m}^{\alpha -1}=\frac{1}{2}\alpha \left( 2-\alpha \right) \Gamma \left(
\frac{\alpha +1}{2}\right)   \label{Wm}
\end{equation}%
(here, as said above, $\varepsilon \equiv 1$ is set). The substitution of
value (\ref{Wm}) in expression (\ref{P2}) with $\sigma =2$ and $\varepsilon
=1$ yields the maximum value of the power, as a function of $\alpha $:%
\begin{equation}
\left( P_{\max }\right) _{\mathrm{VA}}^{\left( \sigma =2,\xi =0\right)
}(\alpha )=\sqrt{\frac{3}{2}\sqrt{3\pi }\left( \alpha -1\right) \left( \frac{%
\alpha }{2}\Gamma \left( \frac{\alpha +1}{2}\right) \right) ^{1/\left(
1-\alpha \right) }\left( 2-\alpha \right) ^{\left( 2-\alpha \right) /\left(
\alpha -1\right) }}.  \label{Pmax}
\end{equation}%
In particular, the limit value of this expression for $\alpha \rightarrow 2$
is
\begin{equation}
\left( P_{\max }\right) _{\mathrm{VA}}^{\left( \sigma =2,\xi =0\right)
}(\alpha =2)=3^{3/4}\approx 2.28.  \label{alpha=2}
\end{equation}%
Naturally, this value exceeds its counterpart (\ref{VA-TS}) corresponding to
$\varepsilon =0$.

Further consideration demonstrates that Eq. (\ref{Wm}) yields values which
monotonously decrease from $W_{m}=1.2$ towards $W_{m}=0$ as LI decreases
from $\alpha =1.5$ to $\alpha =0$. This fact implies that, at $\alpha <1.5$,
$W_{m}$ is not large enough in comparison to the value $\xi =0.05$ adopted
in Eq. (\ref{delta}), hence the simplified VA prediction (\ref{Pmax}),
obtained by setting $\xi =0$ in Eq. (\ref{P}), is not relevant for the
comparison with numerical results at $\alpha <1.5$.

\section{Numerical results for solitons pinned to the attractive defect}

Numerical solutions of Eqs. (\ref{NLS}) and (\ref{u}), with the
delta-function approximated by expression (\ref{delta}), where performed,
chiefly, in the computational domain $-8<x<+8$, discretized by $2^{12}$
points. This domain size is sufficient, as it is much larger than the width
of numerically found solitons, see Figs. \ref{fig3} and \ref{fig6} below.
Stationary solutions were produced by means of the modified square operator
method \cite{Yang}, with zero boundary conditions at $x=\pm 8$. Simulations
of the evolution were run by means of the split-step Fourier method with
longitudinal step $\Delta z=0.0001$.

\subsection{The case of the quintic self-focusing ($\protect\sigma =2,
\protect\alpha <2$)}

\begin{figure}[h]
\centering
\includegraphics[width=3in]{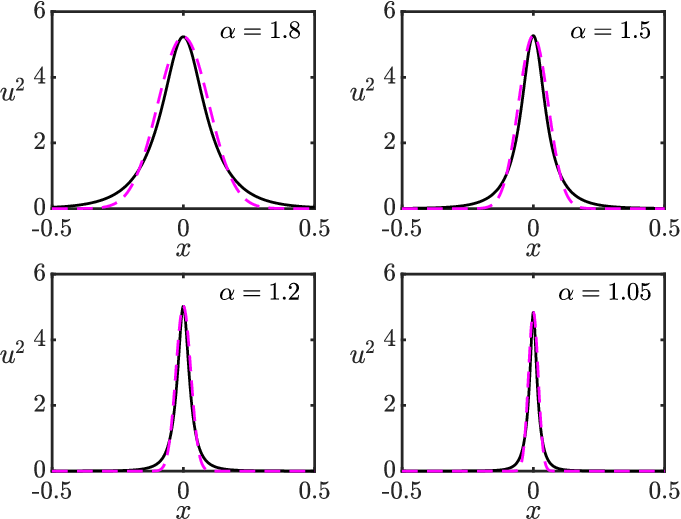}
\caption{Power profiles, $\left( u(x)\right) ^{2}$, of the numerically found
and VA-predicted stable solitons are displayed by continuous black and
dashed magenta lines, respectively, in the case of the quintic self-focusing
($\protect\sigma =2$) and propagation constant $k=10$ in Eq. (\protect\ref{u}
). The corresponding LI values $\protect\alpha $ are indicated in panels.
The solitons are pinned to potential $U(x)=-\tilde{\protect\delta}(x)$, see
Eq. (\protect\ref{delta}). The numerical results are produced as solutions
of Eq. (\protect\ref{u}) with the delta-function approximated as per Eq. (%
\protect\ref{delta}). The VA profiles are based on ansatz (\protect\ref{ans}%
), with parameters obtained from the numerical solutions of Eqs. (\protect
\ref{k(P)}, (\protect\ref{P}) and (\protect\ref{Nans}).}
\label{fig3}
\end{figure}

As mentioned above, the quintic nonlinearity ($\sigma =2$) makes all
solitons unstable in the free space (with $U=0$) for all values of LI $%
\alpha \leq 2$, while the quasi-delta-functional potential $U(x)=-\tilde{%
\delta}(x)$ (see Eq. (\ref{delta})) may stabilize them. Typical examples of
numerically found\emph{\ stable} solitons pinned by the potential are
displayed in Fig. \ref{fig3}, along with their VA-predicted counterparts,
for a fixed propagation constant, $k=10$, and LI values $\alpha =1.8$, $1.5$%
, $1.2$, and $1.05$. It is seen that the VA, which is based on the numerical
solution of the Euler-Lagrange equations (\ref{k(P)}) and (\ref{P}), is
quite accurate in these cases. The decrease of $\alpha $ makes the effective
diffraction weaker, hence the solitons become narrower.

\begin{figure}[h]
\centering
\includegraphics[width=3in]{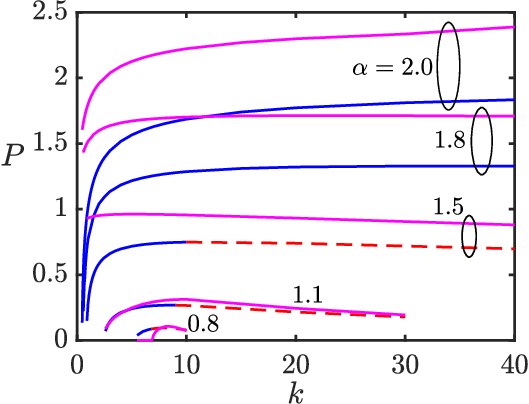}
\caption{The power, $P$, of families of the fractional solitons pinned by
the attractive potential $-\tilde{\protect\delta}(x)$, as produced by the
numerical solution of Eq. (\protect\ref{u}),  vs. the propagation constant, $%
k$, for the quintic nonlinearity ($\protect\sigma =2$) and LI values $%
\protect\alpha $ indicated in the figure. Solid blue and dashed red lines
denote stable and unstable solitons, respectively, the stability being
identified by simulations of the perturbed propagation of the solitons in
the framework of Eq. (\protect\ref{NLS}). The VA-predicted soliton families,
produced by the numerical solution of Eqs. (\protect\ref{k(P)}) and (\protect
\ref{P}), are plotted by solid magenta lines.}
\label{fig1}
\end{figure}

The results for the existence and stability of the pinned solitons in this
model are summarized in Fig. \ref{fig1}, in the form of dependences of the
total power on the propagation constant, $P(k)$, for five LI values, $\alpha
=2.0$, $1,8$, $1.5$, $1.1$, and $0.8$ (the first one, $\alpha =2.0$,
pertains to the NLSE with the usual (non-fractional) diffraction, for which
the full stability of the pinned solitons was demonstrated in Ref. \cite%
{Beijing}). A remarkable fact is that the stabilization remains valid, in a
narrow interval of values of $k$, even for the value of LI as small as $%
\alpha =0.8$, and the simple VA, based on the Gaussian ansatz (\ref{ans}),
is quite accurate for low values of LI (which correspond to strong
fractionality), \textit{viz}., for $\alpha =0.8$ and $1.1$, becoming less
accurate for larger values of $\alpha $. It is also seen that the
instability boundary, located at points with $dP/dk=0$, precisely conforms
to the VK criterion.

\begin{figure}[h]
\centering
\includegraphics[width=3in]{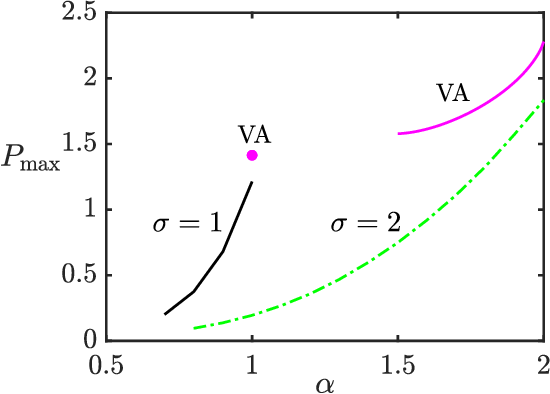}
\caption{The largest power $P_{\max }$, up to which the pinned solitons are
produced by the numerical solution of Eq. (\protect\ref{u}), vs. LI $\protect%
\alpha $. As indicated in the figure,\ the dependences $P_{\max }(\protect%
\alpha )$ are plotted for the cubic and quintic self-focusing ($\protect%
\sigma =1$ and $2$, respectively). For $\protect\sigma =2$, the dependence
predicted by the simplified VA (Eqs. (\protect\ref{Pmax}) and (\protect\ref%
{alpha=2})) is displayed by the magenta line (as explained in the text, it
is irrelevant for $\protect\alpha <1.5$). The magenta dot shows the largest
value of $P$ predicted by the VA for $\protect\sigma =1$, as per Eq. (%
\protect\ref{VA-TS}).}
\label{fig2}
\end{figure}

As mentioned above, both numerical and VA results demonstrate that, for
given LI $\alpha $, the pinned solitons exist below a certain maximum value
of the power, $P<P_{\max }$. The numerically found dependence $P_{\max
}(\alpha )$, produced by Eq. (\ref{u}) with $\sigma =2$, is plotted in Fig. %
\ref{fig2}, along with its simplified-VA counterpart, as given by Eqs. (\ref%
{Pmax}) and (\ref{alpha=2}).

\begin{figure}[h]
\centering\includegraphics[width=3in]{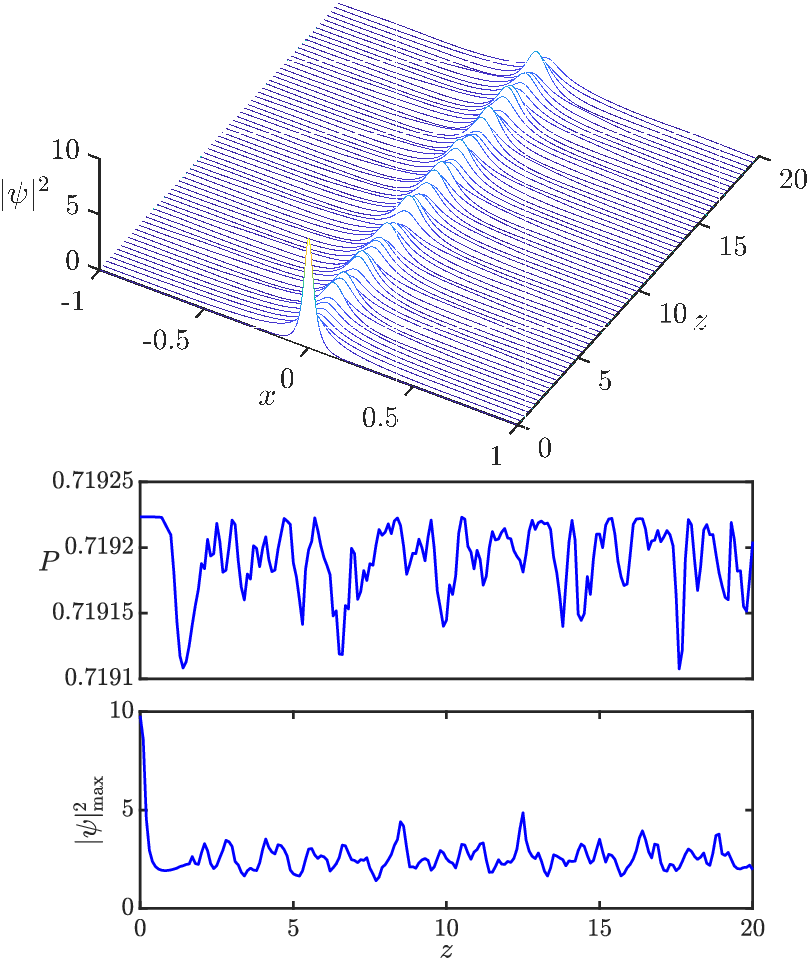}
\caption{{}The top panel: spontaneous transformation of an unstable pinned
soliton, with propagation constant $k=30$, into a breather, produced by
simulations of Eq. (\protect\ref{NLS}) with $\protect\sigma =2$ and $\protect%
\alpha =1.5$. The middle and bottom panels display, severally, the evolution
of the breather's integral power and the evolution of the peak power at $x=0$%
.}
\label{fig14}
\end{figure}

When the pinned solitons are unstable, systematically performed simulations
demonstrate that the instability does not lead to the collapse (blowup of
the solution), which might be expected in the case when the solitons do not
satisfy the VK criterion. Instead, the pinning potential still produces a
stabilizing effect, which transforms the unstable stationary solitons into
breathers featuring somewhat irregular inner oscillations, see a generic
example in Fig. \ref{fig14} (a similar trend was reported in Ref. \cite%
{Beijing} in the model with the non-fractional diffraction, $\alpha =2$, and
supercritical self-attraction, which corresponds to $\sigma >2$ in Eq. (\ref%
{NLS})).

\begin{figure}[tph]
\centering
\includegraphics[width=3in]{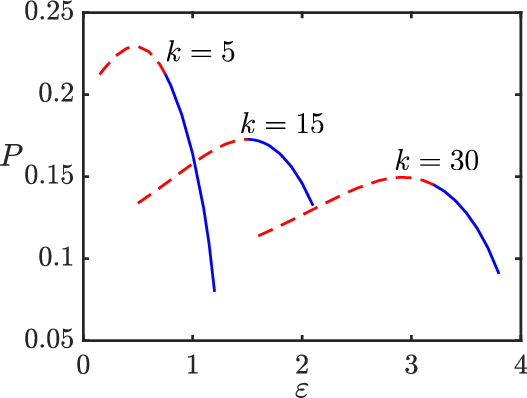}
\caption{Numerically produced $P(\protect\varepsilon )$ dependences for the
solitons pinned to the potential $U(x)=-\protect\varepsilon \tilde{\protect%
\delta}(x)$, for the quintic self-focusing ($\protect\sigma =2$) and $%
\protect\alpha =1$ in Eqs. (\protect\ref{NLS}) and (\protect\ref{u})), with
fixed values of $k$, as indicated in the figure (recall that, unlike the
case of $\protect\alpha \neq 1$, the strength of the attractive $\protect%
\delta $-functional potential cannot be scaled to be $\protect\varepsilon =1$
in the case pf $\protect\alpha =1$). Solid blue and dashed red segments
designate stable and unstable pinned solitons, respectively.}
\label{fig4}
\end{figure}

The above results were obtained, as mentioned above, fixing $\varepsilon =1$
in Eqs. \ref{NLS} and (\ref{u}) by means of rescaling. It was mentioned too
that the rescaling of $\varepsilon $ is not possible in the case of $\alpha
=1$. In the latter case, the existence and stability of the numerically
found pinned solitons are displayed in Fig. \ref{fig4}, by means of
dependences $P(\varepsilon )$ for three fixed values of the propagation
constant, $k=5$, $15$, and $30$. It is observed that, quite naturally, the
stability requires the action of a sufficiently strong pinning potential.

\subsection{The case of the cubic self-focusing ($\protect\sigma =1,\protect%
\alpha \leq 1$)}

\begin{figure}[tph]
\centering
\includegraphics[width=3in]{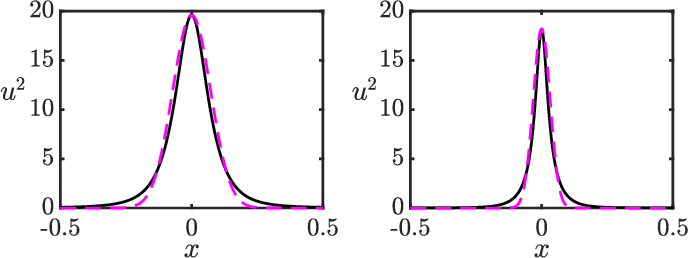}
\caption{Power profiles, $\left( u(x)\right) ^{2}$, of the numerically found
and VA-predicted stable solitons are displayed by continuous black and
dashed magenta lines, respectively, in the case of the cubic self-focusing ($%
\protect\sigma =1$) and propagation constant $k=13$ in Eq. (\protect\ref{u}
). The corresponding LI values $\protect\alpha $ are indicated in panels.
The solitons are pinned to potential $U(x)=-\tilde{\protect\delta}(x)$, see
Eq. (\protect\ref{delta}).}
\label{fig6}
\end{figure}

Examples of stable pinned solitons produced by the numerical solution of Eq.
(\ref{u}) with $\sigma =1$ and two values of LI, $\alpha =1$ and $0.8$, are
plotted in Fig. \ref{fig6}, along with their VA-produced counterparts.
Similar to the examples displayed above for $\sigma =2$ (the quintic
self-focusing) in Fig. \ref{fig3}, it is seen that the VA produces accurate
results.

\begin{figure}[h]
\centering\includegraphics[width=3in]{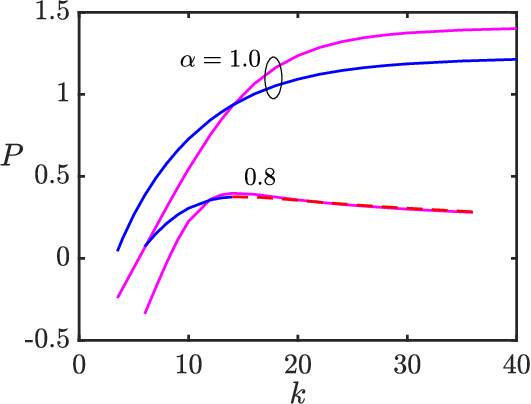}
\caption{The power, $P$, of families of the fractional solitons pinned by
the attractive potential $-\tilde{\protect\delta}(x)$, vs. the propagation
constant, $k$, for the cubic nonlinearity ($\protect\sigma =1$) and LI
values $\protect\alpha $ indicated in the figure. Solid blue and dashed red
lines designate stable and unstable solitons, respectively. The VA-predicted
soliton families, produced by the numerical solution of Eqs. (\protect\ref%
{k(P)}) and (\protect\ref{P}), are plotted by solid magenta lines.}
\label{fig5}
\end{figure}

The results for the existence and stability of the pinned-soliton families
under the action of the cubic self-focusing are summarized, by dint of $P(k)$
curves for two LI values, $\alpha =1$ and $0.8$, in Fig. \ref{fig5}, cf.
Fig. \ref{fig1} for the case of the quintic self-focusing. In particular,
for the critical LI value, $\alpha =1$, the family is completely stable,
similar to the complete stability of the pinned-soliton family for the
critical value $\alpha =2$ in the quintic model \cite{Beijing}. Further, for
$\alpha <1$ the instability boundary is again determined by the VK
criterion, coinciding with the point of $dP/dk=0$. Also similar to the
quintic model, in the cubic one the VA accuracy improves with the decrease
of LI (which is the case of the main interest), making VA quite accurate for
$\alpha =0.8$.

\begin{figure}[h]
\centering\includegraphics[width=3in]{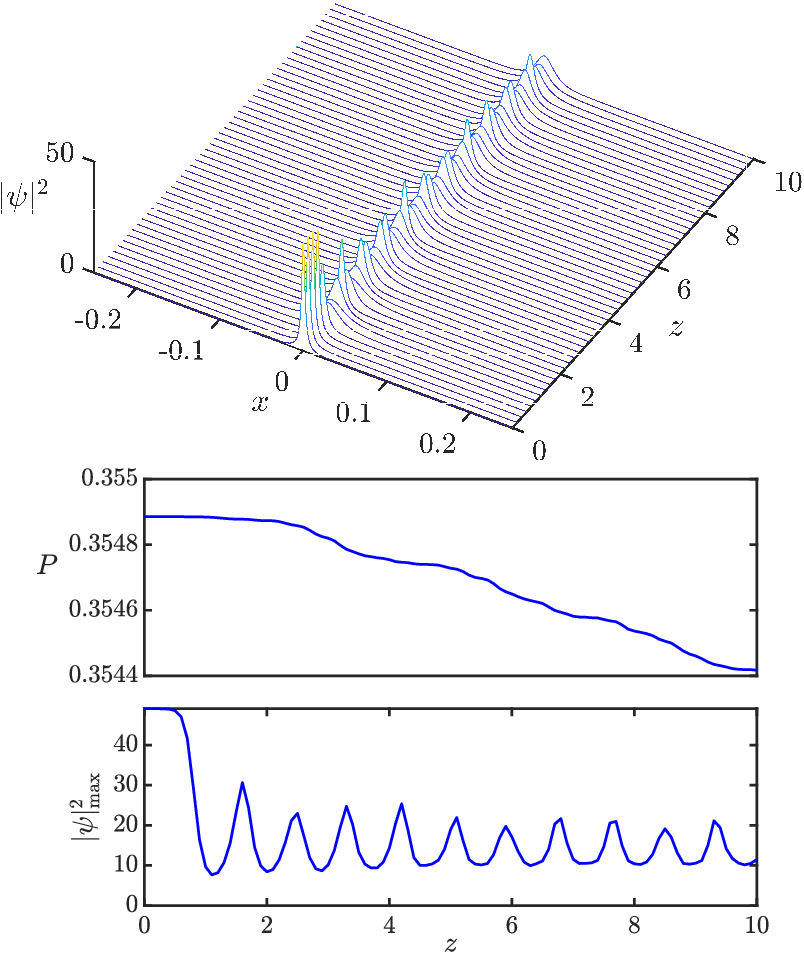}
\caption{{}The top panel: spontaneous transformation of an unstable pinned
soliton, with propagation constant $k=20$, into a breather, produced by
simulations of Eq. (\protect\ref{NLS}) with $\protect\sigma =1$ and $\protect%
\alpha =0.8$. The middle and bottom panels display, severally, extremely
slow decay of the breather's integral power due to very weak emission of
radiation, and the evolution of the peak power, at $x=0$.}
\label{fig15}
\end{figure}

The evolution of unstable pinned solitons under the action of the cubic
self-focusing term is qualitatively similar to that displayed above in Fig. %
\ref{fig14} in the case of the quintic nonlinearity, i.e., spontaneous
rearrangement into robust breathers, as demonstrated by the typical example
which is presented in Fig. \ref{fig15} for $\sigma =1$ and $\alpha =0.8$. On
the other hand, it is seen in the figure that in the present case, unlike
the one presented in Fig. \ref{fig14}, the inner oscillations of the
breather are almost regular.

\begin{figure}[tph]
\centering
\includegraphics[width=3in]{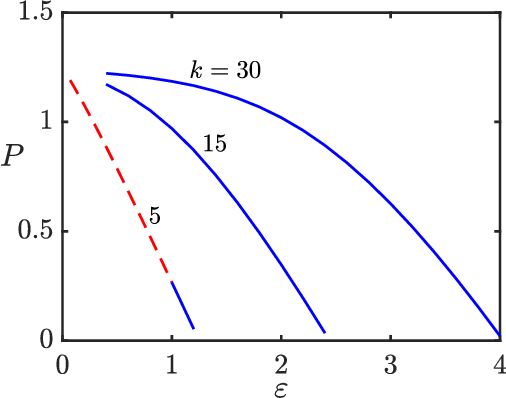}
\caption{Numerically produced $P(\protect\varepsilon )$ dependences for the
solitons pinned to the potential $U(x)=-\protect\varepsilon \tilde{\protect%
\delta}(x)$ in the case of the cubic self-focusing ($\protect\sigma =1$) and
$\protect\alpha =1$ in Eqs. (\protect\ref{NLS}) and (\protect\ref{u})), with
fixed values of $k$, as indicated in the figure (recall that, unlike the
case of $\protect\alpha \neq 1$, the strength of the attractive $\protect%
\delta $-functional potential cannot be scaled to be $\protect\varepsilon =1$
in the case of $\protect\alpha =1$). Solid blue and dashed red segment
designate stable and unstable pinned solitons, respectively.}
\label{fig7}
\end{figure}

As mentioned above, the LI value $\alpha =1$ requires additional
consideration, as in this case the strength of the delta-functional
attractive potential cannot be fixed as $\varepsilon =1$ by means of
rescaling. For this case, properties of the pinned-soliton families are
summarized in Fig. \ref{fig7} by means of $P(\varepsilon )$ curves for three
fixed values of the propagation constant, $k=5$, $15$, and $30$. The
comparison with the similar plot for the model with the quintic
nonlinearity, which is presented above in Fig. \ref{fig4}, demonstrates that
the stability area is larger in the case of the cubic self-focusing. This is
a natural conclusion, as the lower-order nonlinearity implies a weaker drive
for the instability.

\section{The interaction of a free soliton and soliton pair with attractive
and repulsive defects}

\subsection{Collision of a moving soliton with the attractive or repulsive
defect}

In the framework of the interaction of fractional solitons with local
potentials, another relevant problem is collisions of moving solitons (in
fact, tilted ones, in the spatial domain in optics) with attractive and
repulsive defects, in the case of the cubic FNLSE ($\sigma =1$) with $\alpha
>1$, as in this case the solitons are stable and may move in the free space,
even though FNLSE is not a Galilean-invariant equation \cite{Chaos}. To this
end, we address Eq. (\ref{NLS}) with $\sigma =1$, written as%
\begin{equation}
i\frac{\partial \psi }{\partial z}=\frac{1}{2}\,\left( -\frac{\partial ^{2}}{
\partial x^{2}}\right) ^{\alpha /2}\psi -\left\vert \psi \right\vert
^{2}\!\!\psi \mp \tilde{\delta}(x)\psi ,  \label{NLS2}
\end{equation}%
with the bottom sign ($+$) in front of $\tilde{\delta}(x)$ corresponding to
the repulsive defect, unlike the attractive one considered above. The
regularized delta function in Eq. (\ref{NLS2}) is taken as per Eq. (\ref%
{delta}).

\begin{figure}[tph]
\centering
\includegraphics[width=3in]{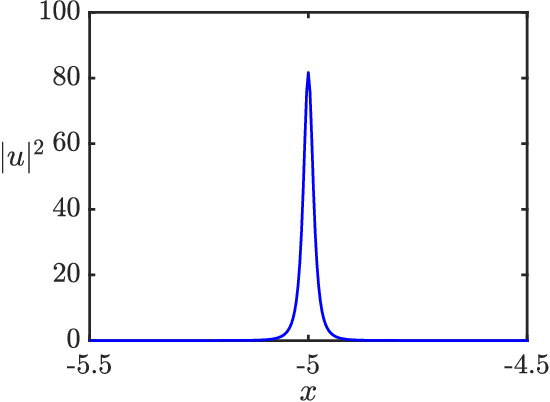}
\caption{A typical profile of a stable narrow "moving" (tilted) soliton,
produced by the numerical solution of Eq. (\protect\ref{u2}) with $\protect%
\alpha =1.2$, $c=0.5$, and $k=30$. The instantaneous coordinate of the
soliton's center is $X_{0}=-5$.}
\label{fig8}
\end{figure}

Far from $x=0$, the moving soliton is sought for as%
\begin{equation}
\psi \left( x.z\right) =e^{ikz}u(X),~X\equiv x-cz,  \label{k2}
\end{equation}%
where $k$ is the propagation constant, and $c$ is the \textquotedblleft
velocity" (actually, it is the tilt of the soliton's trajectory in the
spatial domain), and function $u(X)$ is complex, unlike real $u(x)$ in Eq. (%
\ref{psiphi}). The substitution of ansatz (\ref{k2}) in Eq. (\ref{NLS2})
yields, far from the defect, the stationary equation in the moving reference
frame:%
\begin{equation}
ku+ic\frac{du}{dX}+\frac{1}{2}\,\left( -\frac{d^{2}}{dX^{2}}\right) ^{\alpha
/2}u-|u|^{2}u=0,  \label{u2}
\end{equation}%
cf. Eq. (\ref{u}). Numerical soliton solutions of Eq. (\ref{u2}) were
produced by splitting complex function $u(X)$ into its real and imaginary
parts and solving the corresponding system of coupled real equations. A
typical narrow soliton, with the instantaneous coordinate $X_{0}=5$ of its
center, is presented in Fig. \ref{fig8}.

\begin{figure}[tph]
\centering
\includegraphics[width=3in]{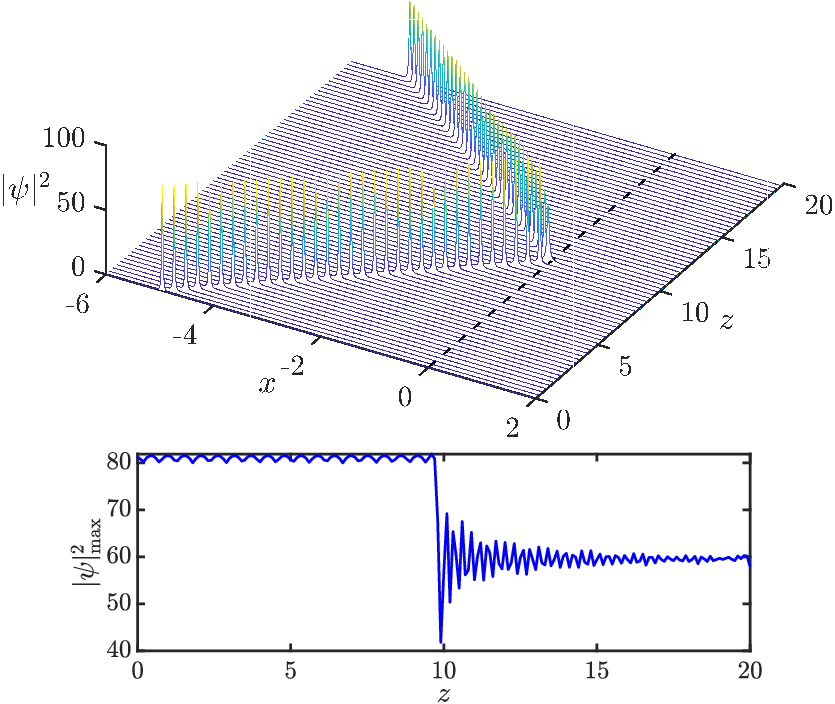}
\caption{The rebound of the incident soliton, with parameters $\protect%
\alpha =1.2$, $k=30$, and $c=0.5$, from the repulsive defect, as produced by
simulations of Eq. (\protect\ref{NLS2}) with sign $+$ in front of the
delta-function. The bottom panel shows the evolution of the soliton's peak
power.}
\label{fig9}
\end{figure}

As shown in Fig. \ref{fig9}, the collision of the incident soliton with the
repulsive defect naturally leads to rebound with velocity $-c$, if $c$ is
smaller than a threshold value, which is $c_{\mathrm{thr}}\approx 0.55$ for
the soliton with $\alpha =1.2$ and $k=30$ in Eq. (\ref{u2}). The evolution
of the soliton's peak power, which is plotted in the bottom panel of Fig. %
\ref{fig9}, demonstrates that the rebound is not fully elastic, as a part of
the power is lost with radiation emitted in the course of the collision.

\begin{figure}[tph]
\centering
\includegraphics[width=3in]{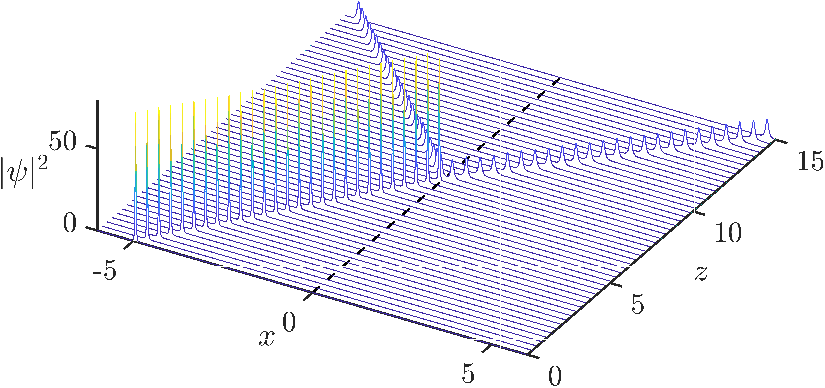}
\caption{The splitting of the soliton colliding with the repulsive defect,
produced by simulations of Eq. (\protect\ref{NLS2}) with sign $+$ in front
of the delta-function. The parameters are $\protect\alpha =1.2$, $k=30$, and
$c=0.625$,}
\label{fig10}
\end{figure}

\begin{figure}[tph]
\centering
\includegraphics[width=3in]{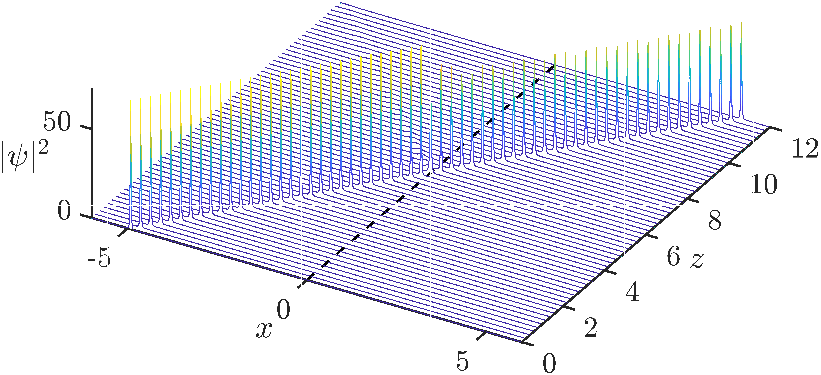}
\caption{The fast soliton passing through the repulsive defect, with
parameters $\protect\alpha =1.2$, $k=30$, and $c=0.82$. The picture is
produced by simulations of Eq. (\protect\ref{NLS2}) with sign $+$ in front
of the delta-function.}
\label{fig11}
\end{figure}

Further, in the interval of the incidence velocity $0.55<c<0.72$ the
collision with the repulsive defect leads to splitting of the incident
soliton into passing and bouncing ones, as shown in Fig. \ref{fig10} for $%
c=0.625$. Finally, at $c>0.75$ the moving soliton passes through the
repulsive defect without a velocity loss. An example of the latter outcome
is presented in Fig. \ref{fig11} for $c=0.82$.

\begin{figure}[tph]
\centering
\includegraphics[width=3in]{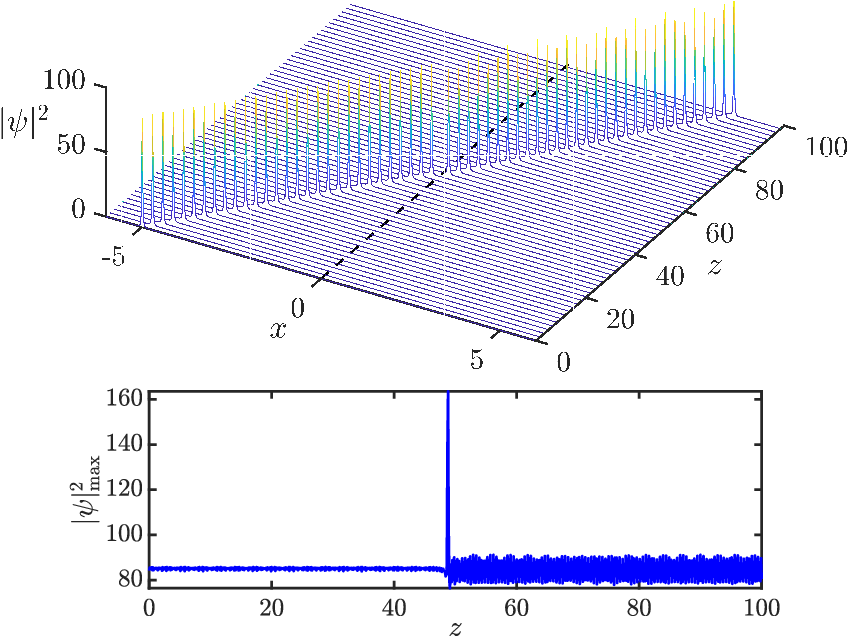}
\caption{The soliton with $\protect\alpha =1.2$, $k=30$, and $c=0.1$ passing
the attractive defect. The bottom panel demonstrates that the collision
leads to excitation of a small-amplitude intrinsic mode in the incident
soliton. The picture is produced by simulations of Eq. (\protect\ref{NLS2})
with sign $-$ in front of the delta-function.}
\label{fig12}
\end{figure}

Finally, the incident soliton with any velocity readily passes the
attractive defect, An example is shown in Fig. \ref{fig12} for a slow
soliton, with velocity $c=0.1$, The evolution of the soliton's peak power,
plotted in the bottom panel, demonstrates that the collision is not fully
elastic, as it causes excitation of an intrinsic mode in the soliton.

\subsection{A two-soliton stationary state pinned to a repulsive defect}

In the case of the normal (non-fractional) diffraction, $\alpha =2$, Eq. (%
\ref{u}) with the general nonlinearity degree, $\sigma $, and repulsive
delta-functional potential ($\varepsilon <0$), admits an exact solution in
the form of a bound state of two solitons with propagation constant $%
k>|\varepsilon |/\sqrt{2}$ and centers placed at points $x=\pm x_{0}$:%
\begin{eqnarray}
\psi _{\mathrm{two-sol}}\left( x,z\right)  &=&\left[ \sqrt{\left( \sigma
+1\right) k}\mathrm{sech}\left( \sqrt{2k}(|x|-x_{0})\right) \right]
^{1/\sigma }e^{ikz},  \label{BS} \\
x_{0} &=&\frac{1}{2\sigma \sqrt{2k}}\ln \left( \frac{\sqrt{2k}+|\varepsilon |%
}{\sqrt{2k}-|\varepsilon |}\right) .  \label{Xi}
\end{eqnarray}%
Roughly speaking, this bound state exists due to the balance between the
repulsive force exerted by the defect on the solitons, and mutual attraction
between the partly overlapping in-phase solitons. However, unlike the
formally similar exact solution for the single soliton pinned to the
attractive defect, which is given by Eqs. (\ref{sol}) and (\ref{xi}), the
two-soliton bound state is unstable, as it is pinned to the repulsive
potential center.

\begin{figure}[tph]
\centering
\includegraphics[width=3in]{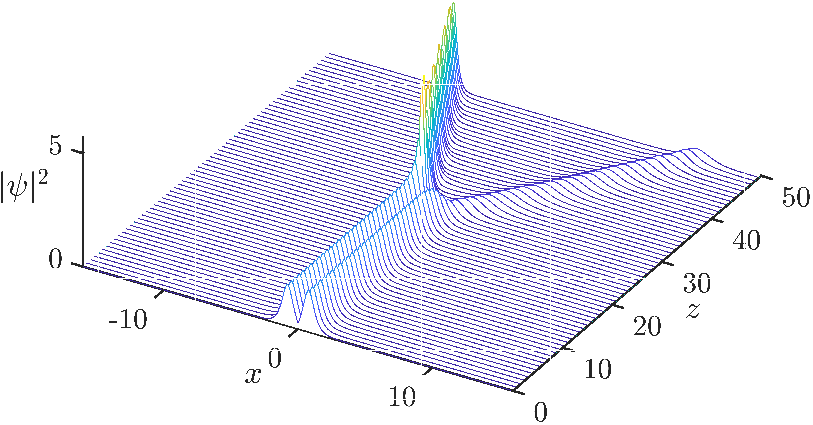}
\caption{The symmetry-breaking evolution of the unstable two-soliton bound
state pinned to the repulsive defect, with parameters $\protect\sigma =1$, $%
\protect\alpha =1.5$ , and $k=0.8$. The picture is produced by simulations
of Eq. (\protect\ref{NLS2}) with sign $+$ in front of the delta-function.}
\label{fig13}
\end{figure}

It may be interesting to consider a similar bound state in the framework of
FNLSE (\ref{NLS2}) with the cubic self-focusing ($\sigma =1$), LI $\alpha >1$%
, and sign $+$ in front of $\tilde{\delta}(x)$. The numerical solution
demonstrates that such states indeed exist, but are unstable against
splitting into free separating solitons, with spontaneous breaking of the
symmetry between them. A typical example is shown in Fig. \ref{fig13} for $%
\alpha =1.5$ and $k=0.8$.

\section{Conclusion}

Currently, much interest is drawn to theoretical investigations and
experimental realizations of nonlinear physical media with effective
fractional diffraction, characterized by the respective value of the LI (L%
\'{e}vy index) $\alpha $. The usual (non-fractional) diffraction corresponds
to $\alpha =2$, while the nontrivial region is $\alpha <2$. In particular,
fractional solitons and other nonlinear modes are promising objects for the
realization in optics. In this context, an important problem is the
stability of fractional solitons, as, in the free 1D space, they are
completely unstable in the case of the cubic self-focusing and $\alpha \leq 1
$. Furthermore, in the case of the quintic self-focusing in the free space,
which also occurs in nonlinear optics, 1D solitons are unstable at all
possible values of \ LI, $\alpha \leq 2$. In this work, we have proposed a
possibility to stabilize the 1D fractional solitons by pinning them to the
attractive defect, which is represented by the regularized delta-functional
potential, in the framework of the corresponding FNLSE (fractional nonlinear
Schr\"{o}dinger equation). This setting is relevant for the creation and
potential use of new species of fractional optical solitons. We have
demonstrated that the fractional solitons with the cubic self-focusing and $%
\alpha =1$ are made completely stable by the pinning to the attractive
defect. The partial stabilization for $\alpha <1$ is identified too.
Similarly, the previously known fact of the complete stabilization of the
usual (non-fractal) solitons by the attractive defect in the 1D medium with
the quintic self-focusing and $\alpha =2$ is extended by the identification
of the stability region for $\alpha <2$. In all cases, the instability
boundary is exactly determined by the VK (Vakhitov-Kolokolov) criterion. In
addition to the systematic numerical analysis, the semi-analytical VA
(variational approximation) is developed. A noteworthy finding is that the
VA becomes quite accurate for lower LI\ values, such as $\alpha =1.1$ and $%
0.8$ for the quintic and cubic self-focusing, respectively, which correspond
to stronger fractionality. Collisions of freely moving stable solitons in
the cubic medium with the repulsive and attractive defects are considered
too, by means of systematic simulations. The increase of the collision
velocity leads to the transition of the outcome of the collisions with the
repulsive defect from rebound to splitting and, eventually, to passage.

As an extension of the work, it may be interesting to consider the interplay
of the fractional diffraction, self-focusing, and a potential cage (cavity)
defined by a pair of repulsive defects. In particular, it may be interesting
to consider shuttle motion of a stable soliton trapped in the cage.

\section*{Acknowledgment}

The work of B.A.M. was supported, in part, by the Israel Science Foundation
through grant No. 1695/22.


\begin{thebibliography}{99}
\bibitem{Za1} A. I. Saichev and G. M. Zaslavsky, Fractional kinetic
equations: Solutions and applications,\textquotedblright\ Chaos \textbf{7},
753--764 (1997).

\bibitem{Za2} G. M. Zaslavsky, Chaos, fractional kinetics, and anomalous
transport, Phys. Rep. \textbf{371}, 461--580 (2002).

\bibitem{Laskin1} N. Laskin, Fractional quantum mechanics and L\'{e}vy path
integrals,\ Phys. Lett. A \textbf{268}, 298--305 (2000).

\bibitem{Stickler} B. A. Stickler, Potential condensed-matter realization of
space-fractional quantum mechanics: The one-dimensional L\'{e}vy crystal,
Phys. Rev. E \textbf{88}, 012120 (2013).

\bibitem{Pinsker} F. Pinsker, W. Bao, Y. Zhang, H. Ohadi, A. Dreismann, and
J. J. Baumberg, Fractional quantum mechanics in polariton condensates with
velocity-dependent mass, Phys. Rev. B \textbf{92}, 195310 (2015).

\bibitem{Laskin-book} N. Laskin, \textit{Fractional Quantum Mechanics}
(World Scientific, Singapore, 2018).

\bibitem{Longhi} S. Longhi, Fractional Schr\"{o}dinger equation in optics,
Opt. Lett. \textbf{40}, 1117--1120 (2015).

\bibitem{Photonics} B. A. Malomed, Optical solitons and vortices in
fractional media: A mini-review of recent results, Photonics \textbf{8}, 353
(2021).

\bibitem{Chaos} B. A. Malomed, Basic fractional nonlinear-wave models and
solitons, Chaos \textbf{34}, 022102 (2024).

\bibitem{Shilong-linear} S. Liu, Y. Zhang, B. A. Malomed, and E. Karimi,
Experimental realisations of the fractional Schr\"{o}dinger equation in the
temporal domain, Nature Comm. \textbf{14}, 222 (2023).

\bibitem{frac-sol} S. Secchi and M. Squassina, Soliton dynamics for
fractional Schr\"{o}dinger equations, Applicable Analysis \textbf{93},
1702--1729 (2014).

\bibitem{LDnonlocal} L. W. Dong, C. M. Huang, and W. Qi, Nonlocal solitons
in fractional dimensions, Opt. Lett. \textbf{44}, 4917-4920 (2019).

\bibitem{LZcoupled} L. Zeng, M. R. Beli\'{c}, D. Mihalache, J. Li, D. Xiang,
X. Zeng, and X. Zhu, Solitons in a coupled system of fractional nonlinear
Schr\"{o}dinger equations. Physica D 456, 133924 (2023).

\bibitem{Shilong-nonlinear} S. Liu, Y. Zhang, S. Virally, E. Karimi, B. A.
Malomed, and D. V. Seletskiy, Observation of the spectral bifurcation in the
fractional nonlinear Schr\"{o}dinger equation, Laser \& Phot. Rev. \textbf{%
2025}, 2401714 (2025).

\bibitem{Martijn} V. T. Hoang, J. Widjaja, Y. L. Qiang, A. F. J. Runge, and
C. M. de Sterke, Observation of fractional evolution in nonlinear optics ,
arXiv:2410.23671.

\bibitem{Mandelbrot} B. B. Mandelbrot, \textit{The Fractal Geometry of
Nature } (W. H. Freeman, New York, 1982).

\bibitem{Riesz} M. Cai and C. P. Li, On Riesz derivative, Fract. Calc. Appl.
Anal. \textbf{22}, 287--301 (2019).

\bibitem{Caputo} M. Caputo, Linear model of dissipation whose Q is almost
frequency independent. II,\textquotedblright\ Geophys. J. Int. \textbf{13},
529--539 (1967).

\bibitem{Vakh} N. G. Vakhitov and A. A. Kolokolov, Stationary solutions of
the wave equation in a medium with nonlinearity saturation, Radiophys.
Quantum Electron. \textbf{16}, 783--789 (1973).

\bibitem{Berge} L. Berg\'{e}, Wave collapse in physics: Principles and
applications to light and plasma waves, Phys. Rep. \textbf{303}, 259--370
(1998).

\bibitem{Fibich} G. Fibich, \textit{The Nonlinear Schr\"{o}dinger Equation:
Singular Solutions and Optical Collapse} (Springer, Heidelberg, 2015).

\bibitem{frac-coll} M. Chen, S. Zeng, D. Lu, W. Hu, and Q. Guo, Optical
solitons, self-focusing, and wave collapse in a space-fractional Schr\"{o}%
dinger equation with a Kerr-type nonlinearity, Phys. Rev. E \textbf{98},
022211 (2018).

\bibitem{AS} F. Kh. Abdullaev and M. Salerno, Gap-Townes solitons and
localized excitations in low-dimensional Bose-Einstein condensates in
optical lattices, Phys. Rev. A \textbf{72}, 033617 (2005).

\bibitem{Beijing} L. Wang, B. A. Malomed, and Z. Yan, Attraction centers and
$\mathcal{PT}$-symmetric delta-functional dipoles in critical and
supercritical self-focusing media, Phys. Rev. E \textbf{99}, 052206 (2019).

\bibitem{Cid1} A. S. Reyna and C. B. de Ara\'{u}jo, Spatial phase modulation
due to quintic and septic nonlinearities in metal colloids, Opt. Exp.
\textbf{22}, 22456 (2014).

\bibitem{Cid2} A. S. Reyna and C. B. de Ara\'{u}jo, High-order optical
nonlinearities in plasmonic nanocomposites -- a review, Advances in Optics
and Photonics. \textbf{9}, 720-774 (2017).

\bibitem{LD1} C. M. Huang and L. W. Dong, Gap solitons in the nonlinear
fractional Schr\"{o}dinger equation with an optical lattice, Opt. Lett.
\textbf{41}, 5636-5639 (2016).

\bibitem{LZ1} L. Zeng and J. Zeng, One-dimensional solitons in fractional
Schr\"{o}dinger equation with a spatially periodical modulated nonlinearity:
nonlinear lattice. Optics Letters \textbf{44}, 2661-2664 (2019).

\bibitem{ZZ} L. Zeng and J. Zeng, Preventing critical collapse of
higher-order solitons by tailoring unconventional optical diffraction and
nonlinearities, Communications Physics \textbf{3}, 26 (2020).

\bibitem{LZ2} X. Zhu, M, R. Beli\'{c}, D. Mihalache, D. Xiang, and L. Zeng,
Two-dimensional gap solitons supported by a
parity-time-symmetric optical lattice with saturable nonlinearity and
fractional-order diffraction, Optics and Laser Technology \textbf{184},
112426 (2025).

\bibitem{Turitsyn} S. K. Turitsyn, J. E. Prilepsky, S. T. Le, S. Wahls, L.
L. Frumin, M. Kamalian, and S. A. Derevyanko, Nonlinear Fourier transform
for optical data processing and transmission: advances and perspectives.
Optica \textbf{4}, 307-322 (2017).

\bibitem{VA} S. I. Muslih, O. P. Agrawal, and D. Baleanu, A Fractional Schr%
\"{o}dinger equation and its solution. Int. J. Theor. Phys. \textbf{49},
1746-1752 (2010).

\bibitem{Yang} J. Yang, \textit{Nonlinear Waves in Integrable and
Nonintegrable Systems} (SIAM, Philadelphia, 2010).
\end{thebibliography}
\end{document}